\documentclass[11pt]{article}

\def\bec{\begin{center}}
\def\enc{\end{center}}

\def\ben{\begin{equation}}
\def\ba{\begin{array}}
\def\bea{\begin{eqnarray}}

\def\een{\end{equation}}
\def\eea{\end{eqnarray}}
\def\ea{\end{array}}
\def\btab{\begin{table}}
\def\btabu{\begin{tabular}}
\def\etab{\end{table}}
\def\etabu{\end{tabular}}
\def\bit{\begin{itemize}}
\def\eit{\end{itemize}}
\def\bef{\begin{figure}[htb]}
\def\befh{\begin{figure}[!h!]}
\def\enf{\end{figure}}

\def\la{\langle}
\def\ra{\rangle}
\def\d{\partial}
\def\a{\alpha}

\def\gb{\beta}
\def\gbL{\beta_L}
\def\gbP{\beta_P}
\def\gU{\Upsilon}
\def\D{\Delta}
\def\de{\delta}
\def\e{\epsilon}

\def\L{\Lambda}

\def\s{\sigma}

\def\b1{{\bf 1}}

\def\bv{\mbox{\boldmath $v$}}
\def\bu{\mbox{\boldmath $u$}}
\def\bn{\mbox{\boldmath $n$}}
\def\bp{\mbox{\boldmath $p$}}
\def\br{\mbox{\boldmath $r$}}

\def\bM{\mbox{\boldmath $M$}}
\def\barM{\bar{M}}
\def\bart{\bar{t}}
\def\barp{\bar{\mbox{\boldmath $p$}}}

\def\nn{\nonumber}
\def\bb{\left(}
\def\eb{\right)}

\def\br{{\bf r}}
\newcommand{\name}{\arabic{section}}
\newcommand{\newsection}[1]{\section{#1}\renewcommand{\theequation}
                              {\name.\arabic{equation}}
                            \setcounter{equation}{0}}
\newcommand{\newsubsection}[1]{\subsection{#1}\renewcommand{\theequation}
                               {\name.\arabic{subsection}.\arabic{equation}}
                               \setcounter{equation}{0}}

\def\rw{\rule[-5mm]{0mm}{12mm}}
\def\rb{\raisebox{3mm}[0pt]}
\usepackage{epsfig,latexsym}

\newcommand{\fm}{{\rm fm}}
\newcommand{\dsp}{\displaystyle}

\begin{document}
\title{Measuring the aspect ratio renormalization of
      anisotropic-lattice gluons }
\author{{\bf M. Alford,}\\
        \\
         Center for Theoretical Physics,\\
         Massachusetts Institute of Technology,\\
         Cambridge, MA 02139, USA,\\
        \\
        {\bf  I.T. Drummond, R.R. Horgan, H. Shanahan, }\\
        \\
        D.A.M.T.P.\\ 
        CMS, Wilberforce Road, \\
        Cambridge, England CB3 0WA\\
        \\
        {\bf M. Peardon,\thanks{Present address: 
             Dept.~of Physics, UCSD,
             La Jolla, California 92093-0319, USA.}
        }\\
        \\
        John von Neumann Institute for Computing and DESY,\\
        Forschungszentrum J\"ulich, D-52425 J\"ulich, Germany.
%       \\
        }
\maketitle
\begin{abstract}
Using tadpole inproved actions we investigate the consistency
between different methods of measuring the aspect ratio renormalization 
of anisotropic-lattice gluons for bare aspect ratios $\chi_0=4,6,10$
and inverse lattice spacing in the range $a_s^{-1}=660-840$ MeV. 
The tadpole corrections to the action, which are established self-consistently, 
are defined for two cases, mean link tadpoles in Landau gauge and gauge 
invariant mean plaquette tadpoles. Parameters in the latter case
exhibited no dependence on the spatial lattice size, $L$, while in 
the former, parameters showed only a weak dependence on $L$ easily 
extrapolated to $L=\infty$~.

The renormalized anisotropy $\chi_R$ was measured using both the torelon 
dispersion relation and the sideways potential method. We found good agreement
between these different approaches. Any discrepancy was at worst $3-4\%$
which is consistent with the effect of lattice artifacts that for the torelon 
we estimate as $O(\a_Sa_s^2/R^2)$ where $R$ is the flux-tube radius.

We also present some new data that suggests that rotational invariance is 
established more accurately for the mean-link action than the plaquette action.

\end{abstract}
\vskip 5 truemm
\begin{center}
keywords: lattice, gauge, anisotropic, QCD
\end{center}
\vfill
DAMTP-1999-59\\
HLRZ1999/49\\
MIT-CTP-2963
\newpage
\newsection{\label{introduction}\bf Introduction}

In principle, an improved action makes it possible to achieve
lattice volumes large enough to overcome finite size effects at a
computational effort low enough to obtain measurements with 
good statistical errors. However typical masses for heavy states may be similar 
to or larger than the inverse lattice spacing on the associated coarse lattice. 
Propagators for such states then decay too fast in lattice units for accurate 
measurement.

The problem can be overcome by tuning couplings associated with the
``time'' direction so that the temporal lattice spacing $a_t$ is
much smaller than the spatial lattice spacing $a_s$~. Refined measurements
are then possible while retaining the computational advantages of the
improved action and coarse spatial lattice.
Such anisotropic actions have already been successfully applied to
the glueball spectrum \cite{MP,MP0}, the spectrum of excitations of the
inter-quark potential \cite{JKM}, heavy hybrids \cite{maea1,drea0},
and the fine structure of the quarkonium spectrum \cite{maea0}.
They are also expected to be a powerful tool in extracting excited-state
signals, obtaining high-momentum form factors,
pushing thermodynamic calculations to higher temperatures,
and in the calculation of transport coefficients.

In this paper, we consider one 
formulation of the anisotropic action for the pure Yang-Mills sector 
\cite{alea,morn}:  
\bea 
\lefteqn{{S}_{n} ~=~} \nn\\
&&-\gb \sum_{x,\,s>s^\prime} \chi_0^{-1} \left\{ 
\frac{5}{3}  \frac{P_{s,s^\prime}}{u_s^4} - 
\frac{1}{12} \frac{R_{ss,s^\prime}}{u_s^6} -
\frac{1}{12} \frac{R_{s^\prime s^\prime,s}}{u_s^6}\right\} - 
\gb \sum_{x,\,s} \chi_0\,\left\{ \frac{4}{3}  
\frac{P_{s,t}}{u_s^2 u_t^2} - \frac{1}{12} \frac{R_{ss,t}}{u_s^4 u_t^2} \right\}~.\nn
\eea
Here $s,s^\prime$ run over spatial directions, $P_{s,s^\prime}$ is a $1\times 1$ 
plaquette and $R_{s^\prime s^\prime,s}$ is a $2\times 1$ rectangle. 
The coefficients of these terms are chosen so the action has no
$O(a_s^2)$ discretization errors in tree-level pertubation theory. 
$\chi_0$ is the anisotropy parameter and is equal to the aspect ratio of the
spatial and temporal lattice spacings, $a_s$ and $a_t$ at tree-level.
At higher orders in the perturbative expansion, this aspect ratio recieves 
quantum corrections, so a renormalized anisotropy determined from a physical 
process, $\chi_R$, differs from $\chi_0$ at $O(\alpha_s)$.
Tadpole improvement (TI) of the perturbative expansion is achieved by 
tuning the input spatial and temporal link parameters $u_s$ and $u_t$ 
for self-consistency at each choice of $\beta$ and $\chi_0$.

Here we report on the accurate determination of the tadpole parameters in the 
plaquette and Landau mean-link formulations. (Non-tadpole-improved actions have been
studied in ref. \cite{buea,saea}.) We compare methods for 
measuring the renormalized anisotropy and discuss the importance of lattice artifacts. 
In order that the anisotropic formulation can be used with confidence, 
physically distinct methods for determining the renormalized anisotropy, 
$\chi_R = a_s/a_t$, should agree. Discrepancies in the results for $\chi_R$ 
give a measure of lattice artifacts and an indication of the importance of 
further improvements. Our results have been used in applications of anisotropic 
lattices to heavy quark hybrid states \cite{drea0}.

In section \ref{tadpoles} we report on the determination of the tadpole parameters; in
section \ref{anisotropy} the renormalized anisotropy measurements are presented, in
section \ref{disc_concl} we discuss the results and present conclusions.  

\newsection{\label{tadpoles}\bf Tadpole Parameters}
In both the plaquette and Landau mean-link formulations, the tadpole parameters are 
determined self-consistently and are defined by
\ben
\ba{lclllclll}
u_s&=& \bb P_{s,s^\prime}\eb^{1/4}&~,~~&u_t&=&1&~~~~~~&\mbox{plaquette},\\
\\
u_s&=& \la U_s \ra_{Landau}&~,&u_t&=&\la U_t \ra_{Landau}&&\mbox{Landau},
\ea
\label{eq:tadpole}
\een
where the Landau gauge is defined by the field configuration which maximizes \cite{lepa}
\ben
F(\{U\})~=~\sum_{x\;\mu}\,{1 \over u_\mu\,a_\mu^2}\mbox{ReTr}\left\{ U_\mu(x)~-~
{1 \over 16u_\mu}U_\mu(x)U_\mu(x+\hat{\mu})\right\}~.
\label{eq:lgf}
\een
with respect to gauge transformations. We denote the gauge coupling in 
the two schemes as $\gbL$ and $\gbP$ respectively.

\newsubsection{\label{plaq_tad}Plaquette Tadpoles}
In the plaquette scheme of Eqn.~(\ref{eq:tadpole}), $u_t=1$ and $u_s$ is determined 
self-consistently. The expectation value of the spatial plaquette is computed for a 
range of input parameters $u_s$ close to and spanning the self-consistent value, 
$u^*_s$. A linear interpolation is sufficient to give an accurate value of $u^*_s$. 
This value is then checked in a further Monte-Carlo simulation. The values of $u^*_s$ 
for a set of couplings $\gbP$ and tree-level anisotropies $\chi_0$ are given in 
Table \ref{tab:plaq_u_s}. For the lattice sizes used in simulation, the plaquette 
expectation value is found to be independent of the volume at the four-significant-figure 
level. An extrapolation to infinite volume is unnecessary.

\btab[htb]
\bec
\btabu{|c|c||c|c|c|}
\hline
$\gbP$     &    $\chi_0$ & $u^*_s$   &$(u^*_s)^4$& $\langle P_{s,s'}\rangle \mbox{ at } u^*_s
$ \\ \hline
 2.1      &    6         & 0.774166  & 0.3592    & 0.359136(72) \\ \hline \hline
 2.3      &    4         & 0.793925  & 0.3973    & 0.397267(52) \\ \hline
 2.3      &    6         & 0.791062  & 0.3916    & 0.391621(73) \\ \hline
 2.3      &    10        & 0.789695  & 0.3889    &0.388650(85) \\ \hline
\etabu
\enc
\caption[]{\label{tab:plaq_u_s}\small
Self-consistent plaquette tadpole parameters.
}
\etab

\newsubsection{\label{Landaugauge} Landau Gauge Fixing}
The maximization of $F(U_\mu^g)$ in Eqn.~(\ref{eq:lgf}), 
where $U_\mu^g=g(x)U_\mu(x)g^{\dag}(x+\mu)$ with respect to a gauge transformation 
$\{g(x)\}$, was carried out using the conjugate-gradient method modified to deal 
with the group structure of the link elements. At each stage of the iteration the
the appropriate conjugate-gradient vector $\{\bv(x)\}$, is computed as a covariant 
derivative \cite{drea1}.
\ben
\bv(x)={\d\over{\d \eta(x)}}F(U_\mu^g)|_{\eta=0}~~,
\een
where $g(x)=e^{\eta(x).T}$ and $T$ are the generators of $SU(3)$~. The vector $\{\bv(x)\}$
has elements lying in the Lie algebra of $SU(3)$~. For each vector a series of 
group elements is constructed in the associated one-parameter subgroup, each element
being a given power of the preceding one:
\ben
g_1~=~\exp(\e\bv)~,~~~~~~g_n~=~(g_{n-1})^p~,~~1 < n \le N~.
\een
The local maximum in the direction $\bv$ is found by evaluating $F(U_\mu^{g_p})$ in
descending order from $p=N$ to $p=1$~. The value of $\e$, at any stage, is reduced
until either the sequence exhibits a maximum or remains constant within a preset 
tolerance. 

For large $\chi_0$ the
terms including $U_t$ dominate the expression for $F$ and near to the maximum relatively small
changes in $F$ correspond to quite large changes in $u_s$. Consequently, to be sure that 
$u_s$ as well as $u_t$ is accurately calculated, the maximum must be found to a
sufficiently high accuracy. The criterion chosen was that 
$\delta F/F \le 2\times 10^{-6}$ where $\delta F$ was the accumulated change in $F$ for three 
consecutive iterations of the conjugate-gradient algorithm for which the individual changes
in $F$ were non-zero. This criterion accounts for the observation that from time to time 
close to the maximum the change in $F$ was zero. We chose $\e = 2\times 10^{-5}$, $p=4$, 
and $N=6$.

\newsubsection{\label{newt_raph} Landau Gauge Fixed Tadpoles}
The self-consistent value of $\bu = (u_s,u_t)$ can be determined by a generalized 
Newton-Rapheson method or by linear interpolation. The latter method was used for 
the mean-link tadpoles and was implemented by choosing four input tadpoles 
$\bu_i,~i=1,\ldots,4$ and measuring the four corresponding output tadpoles 
$\bu_i^\prime,~i=1,4$. A linear map is assumed for the incremental vectors 
$\br_i = (\bu_i-\bu_1),~\rightarrow~\br_i^\prime = (\bu_i^\prime-\bu_1^\prime),~ i=2,3,4$: 
\ben
\br_i~=~\bM\br_i^\prime~,~~~~i=2,3,4~,
\een
where $\bM$ is a $2\times 2$ matrix that is determined from the the measured images of 
$\br_2,\br_3$ and checked for consistency against the measured image of $\br_3$~.  The 
self-consistent tadpole $\bu^*$ is then predicted to be
\ben
\bu^* ~=~\bu_1~+~(1-\bM)^{-1}(\br_1^\prime-\br_1)~.
\een
The self-consistency of $\bu^*$ is then checked computationally. This method 
was found to 
be very reliable and only subject to minor adjustments to account for statistical errors. 

Moreover, a meaningful statistical error can be assigned to $\bu^*$ from the image under
$M^{-1}$ of the statistical error box deduced for $\bu^*$ from the measurements values of the
tadpole parameters. Let the image of $\bu^*$ under the map $M$ be
$\bu^{*\prime}$ with statistical error $\delta\bu^{*\prime}$. The value $\bu^*$ is acceptable 
if in terms of components $\bu^{*\prime}-\delta\bu^{*\prime}~<~\bu^*~<\bu^{*\prime}
+\delta\bu^{*\prime}$, and the error quoted on $\bu^*$ is $\delta\bu^*~
=~M^{-1}\;\delta\bu^{*\prime}$. A typical example is for $\gbL=1.8,~\chi_0=4$ 
on a $6^3\times 24$ lattice where
\ben
\bM^{-1}~=~\left(\ba{rr} 0.341&-0.227\\ -0.033&0.957 \ea\right),~
\delta\bu^{*\prime}~=~\left( \ba{r} 3\times 10^{-4}\\ 3\times 10^{-5} \ea \right)
~\Rightarrow~
\delta\bu^*~=~\left(\ba{r} 1\times 10^{-4}\\ 2\times 10^{-5} \ea \right).
\een
In practice it is found that the error on $u^*_s$ is typically reduced by a 
factor of 3 compared with the statistical error from the simulation  
verifying the self-consistency. 

The simulation for the mean-link tadpoles was done on the Hitachi SR2201 computers 
at the Cambridge High Performance Computing Facility and the Tokyo University 
Computer Centre. The lattices used were $L^3\times T$ with $T=\chi_0L$ and $L=6,8$ 
in all cases except one. The high accuracy demanded by the maximization process was 
very time consuming. Also, many configurations were required to achieve the desired 
statistical accuracy. Consequently, 
only one example with $L=10$ was done as a check on the finite-size scaling ansatz, 
$\bu^*(L)~=~\bu^*(\infty)+A/L^2$~.  Typically, for $10^3\times 40$ about 400 
conjugate-gradient iterations were needed taking about 60 seconds per iteration. In 
the cases $L=6,8,10$ for $\gbL=1.8~\chi_0=4$, the fit to the finite-size scaling 
ansatz was very good as can be seen in Figs. \ref{tad_s} and \ref{tad_t}.  For other 
cases the linear extrapolation was assumed to hold. The values measured for  
$\bu^*(L)$ and the extrapolation to $L=\infty$ are given in table \ref{tad_values}.

\btab[htb]
\bec
\btabu{|c|c||c|c|c|c|c|c|}\hline
&&\multicolumn{5}{c|}{L}\\\cline{3-7}
\rb{$\gbL$}&\rb{$\chi_0$}&4&6&8&10&$\infty$\\\hline
&&0.7165(2)&0.7244(1)&0.7260(1)&0.7266(1)&0.7279(2)\\
\rb{1.8}&\rb{4}&0.98124(3)&0.98201(2)&0.98222(2)&0.98227(2)&0.98243(3)\\\hline
&&0.7115(2)&0.7191(1)&0.7202(1)&--&0.7216(3)\\
\rb{1.8}&\rb{6}&0.99167(3)&0.99194(2)&0.99200(1)&--&0.99208(3)\\\hline
&&--&0.7127(1)&0.7143(1)&--&0.7164(3)\\
\rb{1.7}&\rb{4}&--&0.98105(2)&0.98188(2)&--&0.98295(5)\\\hline
&&--&0.7075(1)&0.7086(1)&--&0.7100(3)\\
\rb{1.7}&\rb{6}&--&0.99149(1)&0.99153(1)&--&0.99158(3)\\\hline
\etabu
\enc
\caption[]{\label{tad_values}\small 
Self-consistent Landau mean-link tadpole parameters for various lattice sizes, $L$, including the
$1/L^2$ extrapolation to $L=\infty$.
}
\etab

\newsection{\label{anisotropy}\bf Anisotropy}

In this section we report on two methods for determining $\chi_R$. The first 
uses the dispersion relation for the torelon \cite{tepe} and the second uses 
the comparison of the potential measured in the fine and coarse direction using 
Wilson loops. 

\newsubsection{\label{torelon}\bf The Torelon}
The lattice considered is $S^2\times L\times T$ with typical values $S=8$ in the $x$ and $y$ 
directions, $L=3,4,5$ 
in the $z$ direction, and $T=50$ in the fine or $t$ direction. The lattice has periodic 
boundary conditions and the torelon is created by a Polyakov line that loops 
around the lattice in the $z$ direction. 
The Polyakov line is associated with a particular point in the $(x,y)$-plane and
is constructed from links which are covariantly smeared in a manner similar 
to APE smearing:
\bea
T(x,y,t)&=&Tr\prod_{z=0}^{z=L-1}W_z(x,y,z,t)~,\nn\\
W_z(x,y,z,t)&=&(1+{l^2D^2\over 4m})^mU_z(x,y,z,t)~,
\eea
where $D$ is the appropriate covariant derivative. Typically, $l=1.5$, $m=10$. The state with
momentum $\bp = (p_x,p_y) = (n_x,n_y)(2\pi/Sa_s)$ is 
\ben
T(\bp,t) = \sum_{x,y}\;T(x,y,t)\;e^{i(p_xx+p_yy)}~.
\een
The torelon propagator 
\ben
G_T(\bp,t) = {1\over T}\sum_{t^\prime=0}^{t^\prime=T-1}\la T(\bp,t^\prime)T^*(\bp,t+t^\prime)\ra~,
\een
is measured for various choices of momentum $\bp$ and a correlated simultaneous fit using
SVD decomposition is made to the relativistic dispersion formula:
\bea
G_T(\bp,t)&=&c(\bp)e^{-E({\bf p})a_t\bart}~,\nn\\
E(\bp)a_t&=&{a_s\sqrt{\bp^2~+~M_T^2}\over \chi_R}~,\label{disp}
\eea
where $M_T$ is the torelon mass and $\bart=t/a_t$. The momenta used were $\bn^2 = n_x^2+n_y^2 = 
0,1,2,4,5$ with $S=8$.  An example of the fit obtained is shown in Fig. \ref{tor_fit}
for $\gbL=1.8,~\chi_0=6$. The fit is over the range $2 \le t \le 12$ and has $\chi^2/N_{\rm df}=0.96$
for 43 d.o.f. The fit is good and gives $(M_Ta_s)^2=1.85(4),~\chi_R=3.61(2)$. In Fig.
\ref{tor_disp} the dispersion curves $E(\bp^2)$ versus $\bn^2$ are plotted for 
$\gbL=1.8,~\chi_0=4,~L=3,4,5$. From both Figs. \ref{tor_fit} and \ref{tor_disp} it is clear
that rotational invariance is established in the coarse $xy$ directions.  The
good simultaneous fit to the torelon propagators in all cases shows that the 
continuum dispersion is well satisfied for momenta at any angle to the coordinate axes.

\btab[htb]
\bec
\btabu{|c|c||c|c|c|c|c|c|}\hline
&&\multicolumn{6}{c|}{L}\\\cline{3-8}
$\gbL$&$\chi_0$&\multicolumn{2}{c|}{3}&\multicolumn{2}{c|}{4}&\multicolumn{2}{c|}{5}\\\cline{3-8}
&&$\chi_R$&$(M_Ta_s)^2$&$\chi_R$&$(M_Ta_s)^2$&$\chi_R$&$(M_Ta_s)^2$\\\hline
1.8&4&3.61(1)&0.683(1)&3.61(3)&1.85(4)&3.57(5)&3.39(13)\\\hline
1.8&6&5.32(2)&0.502(1)&5.28(3)&1.49(3)&5.28(5)&2.84(9)\\\hline
1.7&4&3.56(2)&1.16(2)&3.56(4)&2.86(9)&3.66(9)&5.51(36)\\\hline
1.7&6&5.28(2)&0.89(2)&5.28(5)&2.36(6)&5.26(6)&4.25(13)\\\hline
\etabu
\enc
\caption[]{\label{ani_values}\small
Renormalized anisotropies and dimensionless torelon masses for various values of $\gbL$ and $\chi_0$
from a simultaneous fit on $8^2\times L\times 50$ lattice.
}
\etab
In table \ref{ani_values} we give the full set of results for the torelon anisotropies we have
measured. There is no noticeable dependence within errors of $\chi_R$ on $L$ and the dependence
on $\gb$ is small for the two values of $\gb$ used.

From the values $M_T(La_s)$ we can determine the string tension, $\s$ in units of $a_s^{-2}$. We assume 
that $M_T(La_s) = \s(La_s)La_s$ where $\s(La_s)$ is the string tension modified by finite-size corrections 
\cite{lues,ispa}: $\s(La_s) = \s + D/(La_s)^2$. A plot of $\s(La_s)a_s^2$ versus $1/L^2$ is shown 
in Fig. \ref{string_ten} with a linear fit which has $\chi^2/N_{\rm df}=0.05$ with 
$\s a_s^2 = 0.394(3)$.
In another calculation we have determined the absolute value of $a_t$ in $MeV$ by computing the
splitting $\D M_{PS}=M_\gU(1P)-M_\gU(1S)$ for the $\gU$ meson system using $O(mv^6)$ NRQCD \cite{caea}. 
In table \ref{str_ten} we give $a_t^{-1},~a_s^{-1},~\s a_s^2$, the coefficient $D$, 
and $\sqrt{\s}/\D M_{PS}$. This latter ratio is experimentally close to unity but in ref. \cite{caea} 
using isotropic $\gb=6.2$ UKQCD  configurations this ratio was found to be about 1.25 which, 
from table \ref{str_ten}, agrees with our finding except possibly for the $(\gb=1.7, \chi_0)$ lattice 
which has the most coarse lattice spacings. It is generally accepted that the discrepancy 
is due to quenching and the significant outcome is that our results from the anisotropic lattice 
agree well with those of a spatially finer isotropic lattice of $a=0.06$fm, indicating that we 
have correctly reproduced the physics expected for this comparison on lattices with $a_s = 0.25-0.30$fm.
A naive estimate for the coefficient $D$ is obtained by calculating the contribution to $M(L)$ from the
zero-point fluctuations of the torelon regarded as a periodic flux tube. The outcome is 
$D=-\pi/3 \sim -1.05$ which is compatible with our fit of $D \sim -1.35(5)$. The same calculation
predicts that there is no $L$-independent constant contribution to $M(L)$ and this is verified by our 
fits.

\btab[htb]
\bec
\btabu{|c|c|c|c|c|c|c|c|}\hline
$\gbL$&$\chi_0$&$M_b$&$a_t^{-1}$MeV&$a_s^{-1}$MeV&$\s a_s^2$&$\sqrt{\s}/\D M_{PS}$&D\\\hline
1.8&4&5.32&2876(75)&797(21)&0.422(7)&$\sim$1.18(3)&-1.31(6)\\\hline
1.8&6&4.75&4503(45)&839(9)&0.394(3)&1.19(2)&-1.42(3)\\\hline
1.7&4&5.97&2353(38)&661(11)&0.513(12)&1.08(2)&-1.39(13)\\\hline
1.7&6&5.56&4112(150)&779(28)&0.469(8)&$$1.21(5)&-1.38(8)\\\hline
\etabu
\enc
\caption[]{\label{str_ten}\small
The inverse lattice spacings for mean-link tadpole improved actions from the $O(mv^6)$ NRQCD 
measurement of $\D M_{PS}=M_\gU(1P)-M_\gU(1S)$ to determine $a_t^{-1}$ and using the torelon 
anisotropy to infer $a_s^{-1}$. The string tension is found from the $L-$dependence of the 
torelon mass using finite-size scaling ansatz $\s(La_s) = \s + D/(La_s)^2$.
}
\etab

The torelon dispersion relation was also studied on the plaquette-tuned
parameters. For these simulations, lattices of extent $(8 \times 10) \times 4
\times N_t $ were used, to enable us to investigate a wider range of momenta
combinations. At momenta close to the cut-off, we expect the
discretization errors to be $O(a_s^4 p^4,\alpha_s a_s^2 p^2)$, thus in order 
not to
contaminate our determination of $\chi_R$ with large discretization errors, we
first determined the range of momenta over which a good correlated fit to the 
continuum
dispersion relation could be made. The data for different momentum ranges from 
simulations at $\gbP=2.1, \chi_0=6$ were tested and these
results are shown in Table~\ref{tab:plaqtor-momenta}. Both the measured 
anisotropy renormalizations and the torelon rest energies determined are 
all consistent within statistical precision up to the highest momentum 
measured. For the subsequent computations of the anisotropy, presented in 
Table~\ref{tab:plaq-ani-values}, the largest 
momentum used was the (1,1) data (shown in bold in 
Table~\ref{tab:plaqtor-momenta}).

\btab[htb]
\bec
\btabu{|c||c|c|c|}\hline
Maximum    & $\chi_R/\chi_0$ & $a_t M_T$ & $\chi^2/N_{\rm df}$ \\
$(p_x,p_y)$&                 &           &                     \\
\hline
(1,1)      &{\bf 0.949(25) } & 0.2884(17)& 1.19 \\
(0,2)      &     0.972(23)   & 0.2886(16)& 0.99 \\
(1,2)      &     0.968(17)   & 0.2885(15)& 0.85 \\
(2,1)      &     0.976(14)   & 0.2885(15)& 0.89 \\
(2,2)      &     0.970(11)   & 0.2883(14)& 0.96 \\
\hline
\etabu
\caption[]{\label{tab:plaqtor-momenta} Dependence of the renormalized 
anisotropies and torelon masses on the highest momentum used in the fit to
\protect{Eqn.~(\ref{disp})} at $\gbP=2.1, \chi_0=6$.}
\enc
\etab

\btab[htb]
\bec
\btabu{|c|c||c|c|}\hline
$\gbP$ & $\chi_0$ & $\chi_R/\chi_0$ & $\chi^2/N_{\rm df}$ \\
\hline
2.3    & 4        & 0.941(13) & 1.04 \\
2.3    & 6        & 0.954(23) & 0.74 \\
2.3    & 10       & 0.940(13) & 1.41 \\
\hline
2.1    & 6        & 0.949(25) & 1.19 \\
\hline
\etabu
\caption[]{\label{tab:plaq-ani-values}\small
Renormalized anisotropies for various values of
 $\gbP$ and $\chi_0$ using plaquette-tuned mean-link parameters.
}
\enc
\etab

\newsubsection{\label{potential}\bf The Sideways Potential}

% MGA:
In the sideways potential method \cite{buea,klas},
a coarse direction, $z$, on the anisotropic lattice is chosen to be the time direction.
There are then two types of spacelike direction, coarse and fine. 
The measurement of the potential between static quarks with a separation lying 
in the plane of coarse links is compared with the measurement of 
the potential when the separation lies along the line of fine links.
The demand that the two measurements yield the same function of {\it physical} distance 
determines the renormalized anisotropy. Points in the coarse-coarse plane are
denoted by $\vec x=(x,y)$ and points in the fine direction by $t$ where $x,y,z,t$ are integers.

We measure appropriate spatial Wilson loops $W_{ss}(\vec x,z)$ and also loops 
using the fine direction $W_{ts}(t,z)$.  
We define
\begin{equation}
V_s(\vec x,z) = \log \left(\frac{W_{ss}(\vec x,z)}{W_{ss}(\vec x,z+1)}\right), \qquad
V_t(t,z) = \log \left(\frac{W_{ts}(t,z)}{W_{ts}(t,z+1)}\right).\label{potentials_ma}
\end{equation}
As $z\to\infty$, $V_s(\vec x,z)\to V_s(|\vec x|)$,  and $V_t(t,z)\to V_t(t)$ 
where $V_s(|\vec x|)$ and $V_t(t)$ are the two versions of the interquark potential.
For a physical distance $r$ we have $|\vec x|a_s=ta_t=r$~.
We therefore estimate the renormalized anisotropy $\chi_R$ by tuning it so that
$V_s(|\vec x|) = V_t(t/\chi_R)$, where the right side is evaluated by means of linear 
interpolation between the values measured at integral $t$. It is implicit in the
method that there is effective rotational invariance in the $\vec x$-plane. 
We find that if we exclude potentials at the smallest distances,
$|\vec x| = 1,\sqrt{2}$, then the values of $\chi_R$ associated with different
directions in the $\vec x$-plane generally agree within errors. The agreement is
particularly good if the links are smeared in an appropriate manner. 
The results for the renormalized anisotropy, $\chi_R$, for both mean-link 
and plaquette schemes are shown in table {\ref{side_anis}}.

An alternative approach to making the comparison is to fit the measured potentials 
with the forms
\bea
a_s V_s(\vec x) = a_s V_0 + \sigma a_s^2   \; x + \frac{e}{x}  \nonumber \\ 
a_s V_t(t)      = a_s V_0 + \sigma a_s a_t \; t + \frac{a_s e}{a_t t}~.
\label{potentials_mp}
\eea
The renormalized anisotropy, $\chi_R$, is then determined from the ratio of the coefficients
of the linear terms in the two cases. 
$\chi_R$ can in principle be determined from the ratio of the coefficients of 
the coulombic terms however such an estimate depends on short distance effects 
and is inherently more sensitive to discretization errors. 

This approach was tested on an $8^3 \times 48$ lattice at $\gbP=2.3, \chi_0=6$.
Prior to measurement, the lattice is blocked to an $8^4$ volume by thinning the
time-slices of spatial links and replacing the temporal links with the product 
of six fine links connecting the appropriate time-slices. 
The $t$ and $z$ axes are then interchanged, since as before we want to use a
coarse direction for Euclidean decay. Then a standard APE-smearing algorithm 
is applied to links on the new time slices. These
degrees of freedom are then used to measure the potential in the $(x,y)$ and 
$z$ axes in the standard fashion. Fig. \ref{fig:plaq-sideways} shows the
results of this simulation. The best fits of this data to 
Eqn.~(\ref{potentials_mp}) for a range of static-source separations were 
computed and the string tension results are shown in Table~\ref{tab:pot-fits}. 
In Fit C, the coulomb term was poorly resolved and was thus fixed to zero. The 
use of these different ranges allowed us to investigate the systematic 
uncertainty and discretization errors in determining $\chi_R$ from this method. 
The three results are consistent within $2\%$. The anisotropy measurement from 
the full range of separations (Fit A, shown in bold in Table~\ref{tab:pot-fits})
is in good agreement with the torelon dispersion result of 
Table~\ref{tab:plaq-ani-values}.

\btab
\bec
\btabu{|c|c|c||c|c|c|}\hline
\multicolumn{3}{|c||}{mean-link }&\multicolumn{3}{|c|}{plaquette }\\\hline
$~\gb~$&$~\chi_0~$&$~\chi_R~$&$~\gb~$&$~\chi_0~$&$~\chi_R~$\\\hline
1.8&4&3.67(2)&2.1&6&5.80(5)\\\hline
1.8&6&5.48(3)&2.3&4&3.95(3)\\\hline
1.7&4&3.67(3)&2.3&6&5.93(5)\\\hline
1.7&6&5.40(5)&2.3&10&10.00(20)\\\hline
\etabu
\caption[]{\label{side_anis}\small Results for the renormalized
anisotropy, $\chi_R$, using the sideways potential method for
both mean-link and plaquette tadpole-improved schemes}
\enc
\etab

\btab[htb]
\bec
\btabu{|c|c|c|c|c|c|c|}
\hline
Fit & Separations & $\sigma a_s^2$     &$\chi^2/N_{\rm df}$ &
                    $6 \sigma a_s a_t$ &$\chi^2/N_{\rm df}$ & $\chi_R/\chi_0$\\ 
\hline
  A  &     1-7    & 0.3293(33) & 1.13 & 0.3492(26) & 1.21 & {\bf 0.9430(12)} \\
  B  &     1-4    & 0.3315(60) & 0.85 & 0.3526(64) & 1.14 & 0.9402(24) \\
  C  &     3-7    & 0.3506(32) & 0.70 & 0.3664(34) & 1.11 & 0.9569(12) \\ 
\hline
\etabu
\caption[]{\label{tab:pot-fits} Results for the renormalized anisotropy, 
$\chi_R$ from fits to the static potential. In fit C, the Coulomb term was set 
to 0.}
\enc
\etab

Both the above approaches yielded results reasonably consistent with each other 
and with the toleron results at the 3\% level. Discrepancies can easily be explained 
by the presence of discretization errors.
Because the flux tube generating $V_s(|\vec x|)$ has a coarse-fine cross-section
and that generating $V_t(t)$ a coarse-coarse cross-section we anticipate, as in the case
of the toleron (eq.(\ref{torerr}), that some discretization error in $\chi_R$ 
that is $O((\a_sa_s/R)^2)$ will remain.  Since $a_s\sim 0.3~\fm$, $R\sim 0.7$ to $1~\fm$, 
and $\a_s\sim 0.3$, we expect discretization errors to be around 5\%.

Because we are concerned to make the long distance behavior consistent in  
both the fine and coarse directions it is advantageous to use Wilson loops of 
the largest possible spatial extent. However, in practice, the statistical errors 
on large Wilson loops grow exponentially with separation. In order to achieve 
acceptable errors we were restricted to using loops of size 2 or 3 in coarse lattice 
units. This is to be compared with the torelon, where the practical 
size is 3 or 4 lattice units. 

\newsection{\label{disc_err} \bf Discretization Errors}

One way of viewing discretization errors is to regard them as arising
from the absence of the correction term in the lattice action that yields
continuum results.
The corresponding term in the action density can be presumed to be 
a local (redundant) operator. The locality of the operator implies that
while there may be a correction to the mass per unit length of the torelon,
there is no correction that directly alters the asymptotic proportionality
of the torelon mass and its length. The appropriate length scale against which
to measure lattice artifacts therefore is the radius $R$ of the toleron cross-section
or equivalently, $\Lambda_{QCD}^{-1}$ or $\sqrt{\sigma}$ where $\sigma$ is the string tension.  
We expect therefore that the $O(a_s^2)$ errors in $\chi_R$ are proportional to $(a_s/R)^2$~. 

A persuasive plausibility argument for this behavior is as follows. The torelon state 
of momentum $\bp$ and mass $M_T$ is an eigenstate of the 
transfer matrix $T$ with an eigenfunctional labelled by $\bp$ and  $M_T$ and eigenvalue 
$\exp(-E(\bp^2)a_t/\chi_R)$. Suppose an operator is added to the action which corrects 
for the $O(\a_sa_s^2)$ errors. This operator will be local on the scale of the lattice 
and the effect on $E(\bp^2)$ can be estimated using first-order perturbation theory: 
\ben
\delta {a_sE(\bp^2) \over \chi_R}~=~\a_sf(\barp^2,\barM_T)~,
\een
where $\barp^2 = \bp^2a_s^2$, $\barM_T = M_Ta_s$, and $f(\barp^2,\barM_T)$ is the dimensionless 
matrix element of the added operator which is proportional to $a_s^2$ by construction. 
The question is what scale balances the dimension of $a_s^2$? We find
\bea
-{\delta \chi_R \over \chi_R^2}&=&\a_s\phi^\prime(0,\barM_T)~,\label{delta_chi}\\ 
{\delta \barM_T^2  \over 2\chi_R}&=&\a_s(\phi(0,\barM_T)-\barM_T^2\phi^\prime(0,\barM_T))~,
\label{delta_m}
\eea
where $\phi(\barp^2,\barM) = a_sE(\bp^2)f(\barp^2,\barM_T)$ and $\phi^\prime 
= \d\phi/\d(\barp^2)$. It must be that the change in the action by a local operator 
corresponds to a change in the string tension, $\s$. Hence, since $M_T = \s L$, it 
follows that $\delta\barM_T^2 \propto \barM_T^2$ as this is the only parameter depending 
on $L$, and from eqn. (\ref{delta_m}) this implies that
\ben
\phi(0,\barM_T)~\sim~\barM_T^2~,~~~~\phi^\prime(0,\barM_T)
                               ~\sim~\barM_T\mbox{-independent constant}~.
\een
It is conceivable that an accidental cancellation between the two terms in eqn.
(\ref{delta_m})
would allow different behavior to be inferred for $\phi$ and $\phi^\prime$ but these results
must hold true for all possible local perturbations, not just the particular one that 
eliminates  $O(\a_sa_s^2)$ errors. We consider this kind of cancellation to be unlikely. 
Substituting this behavior for $\phi^\prime$ into Eqn.~(\ref{delta_chi}) we find that
\ben
\label{torerr}
{\delta \chi_R \over \chi_R^2}~\sim~C{a_s^2\over R^2}~,
\een
where $R$ is a typical length scale associated with the torelon which cannot be $M_T^{-1}$, 
for example, the flux tube radius. Alternatively, $R^{-1}$ can be taken to be 
$\sqrt{\s}$, $\L_{QCD}$.  

\newsection{\label{disc_concl} \bf Conclusions}

In this paper we have investigated various methods for measuring the
renormalized aspect ratio for pure QCD on an anisotropic lattice. In
the main we used a tadpole-improved action with the Landau gauge mean-field 
definition for the tadpole parameters, but we also included results 
for the action with tadpoles defined by the mean plaquette. The object of the 
investigation was to assess the consistency of different methods in order to judge the
effectiveness of the improvement scheme. 
In principle, measurements of the anisotropy from different 
physical probes should agree close to the Euclidean-symmetric continuum limit.
The bare anisotropy $\chi_0$ is renormalized by the effect of operators which are
irrelevant in the neighborhood of the fixed point controlling 
the continuum limit. From a renormalization group (RG) view point the location of 
this fixed point is ambiguous up to redefinitions of the RG transformation
used to locate it. This ambiguity is due to redundant operators
\cite{wegn,fira,luwe} which have no effect on physical observables of the 
continuum theory. Consequently, continuum actions with different $\chi_R$ can differ only by 
redundant operators since they must correspond to the same continuum physics. 
Thus, while $\chi_R$ can be changed by tuning $\chi_0$, an action with no lattice 
artifacts must give rotationally invariant physical results and, consequently, any physical 
method for measuring $\chi_R$, such as the ratio of two physical observables, 
must give the same answer.  In as much as this is not the case the differences will 
give an estimate for the effect of lattice artifacts and the necessity for improvement.

The spatial lattice spacing $a_s$ was measured in a separate NRQCD simulation by
fitting the $1P-1S$ mass-splitting $\D M_{PS}$ for bottomonium. 
This estimate can be compared with the string tension, deduced from a 
$D/L^2$ extrapolation to $\infty$ of the torelon mass. The coefficient $D$ was 
found to be $\sim -1.38$
which is in tolerable agreement with $-\pi/3$ predicted from analysis of 
flux-tube fluctuations \cite{ispa}. The ratio $\sqrt{\s}/\D M_{PS}$ is $\sim 1.2$ in agreement with
earlier NRQCD analyses \cite{caea0,caea}. The departure from the experimentally 
observed value of $\sim 1$ is attributed to quenching.
The values for both the mean-link and plaquette tadpoles were found self-consistently.
This was very resource intensive in the mean-field case since it required a very 
accurate gauge fixing to Landau gauge and the self-consistent iteration is in
the 2D space of $(u_s,u_t)$ which requires additional effort. Also, for given $\chi_0$ the 
tadpoles showed an $L$ dependence and hence required an extrapolation to $L=\infty$. This
in turn requires accurate data for a good fit. In contrast, the plaquette tadpole
was easily found and there was no discernible $L$-dependence.

The torelon method deduces $\chi_R$ from a fully correlated fit
to the dispersion relation. Statistical errors are produced by the fit.
The sideways potential approach requires some method for matching 
the coarse and fine potentials and the error analysis is more 
complex because of both systematic and statistical errors and because the signal
rapidly decreases as the loop size increases. Two methods were used to extract $\chi_R$.
The first compared the loop predictions for the coarse and fine potentials 
(\ref{potentials_ma}) and deduced $\chi_R$ from the rescaling of $t$ needed for them
to agree. This method was applied to actions with mean-field improvement. Typical loops 
were of edge length 2-3 in units of the coarse spacing $a_s$. There is a clear difference
in results compared with the torelon computation of about $3-4\%$. 
In the second approach the fine direction is first blocked by a scaling of $\chi_0$ to 
give a lattice that is approximately isotropic and then the potential
is fitted to a standard form (\ref{potentials_mp}) and $\chi_R$ deduced by 
requiring that the linear slopes agree. The latter method has the advantage
that it depends only on the long range structure of the potential and excludes the
short range coulomb part. However, it does require large computing resources to
extract a reliable signal at large separations, in this case up to 7 lattice units.
This technique was applied mainly to actions with plaquette improvement.
Although less extensively investigated, the results are consistent with the 
torelon computation. 

The observed discrepancies can be easily attributed to residual lattice artifact 
effects. For the torelon we expect $\de \chi_R/\chi_R \sim C\a_s a_s^2/R^2$ which can
be sizeable although $C$ is unknown. However, it should also be remarked that
neither the torelon nor the second sideways potential method include the coulomb
part of the static potential. Because of its short-range nature this part will be
the most sensitive to the effect of lattice artifacts. In this case agreement 
of prediction for $\chi_R$ is good. This may merely indicate that $\chi_R$ is being
deduced from the properties of the flux tube in both cases and so they should
agree in that artifact effects are the same. However, it is encouraging that 
different methods based on the long-range properties of the action are consistent and
it is worth noting that the torelon flux tube was 3--5 lattice units in length whereas
it was necessary to extend the potential method to 7 lattice units. Simulation times
are correspondingly reduced for the torelon.  Unfortunately, it was not possible to 
apply the second method to data gathered using the mean-field action because of 
limitations in resources.

\begin{figure}[htb]
\begin{center}
\epsfxsize=10cm
\hspace*{0in}
\epsffile{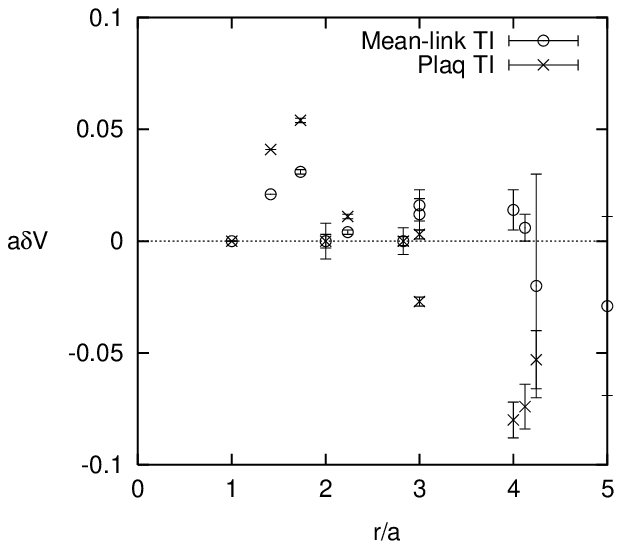}
\end{center}
\caption{
Deviation of static potential from fit, $a\approx 0.29~\fm$.
Both sets of data are fitted to $V(r)= b + c/r + \sigma r$, using
the points $r=1,2,\sqrt{8}$, which therefore show no deviation.
For the remaining points, it is clear that mean-link TI (circles)
shows smaller rotational symmetry violations than plaquette TI.
See also table \ref{tab:rotl}.
}
\label{fig:potl}
\end{figure}

We do not find any evidence that the mean-link tuned action is superior to the 
plaquette tuned action in suppressing lattice artifacts, and it should be noted 
that the computation required to determine the mean-link tadpoles
for the former case is very time consuming. However, there is other evidence
that the mean-link action is superior. It is known to give smaller scaling 
errors in the NRQCD charmonium hyperfine splitting \cite{shtr0,shtr1}, 
better hadron mass scaling \cite{alea}, and to give a clover
coefficient for SW fermions (with the Wilson gauge action) that agrees more 
closely \cite{lepa} with the non-perturbatively determined value 
\cite{somm,luea}. Using our configurations
we present evidence that it also gives better rotational invariance for
the static quark potential. In Fig.~\ref{fig:potl} we show the
deviation of the potential from a fit to the standard form $V(r)= b +
c/r + \sigma r$.  Comparing the mean-link TI data with the plaquette
TI data, it is clear that mean-link TI gives smaller deviations from
the rotationally invariant fit.
In table \ref{tab:rotl} we present some of the
features of Fig.~\ref{fig:potl} in a quantitative form.
As expected, mean-link TI gives better rotational invariance at $r=\sqrt{3}$,
and it causes the two $r=3$ potentials ($(2,2,1)$ vs $(3,0,0)$) to
agree better.

\btab
\bec
\def\bigst{\rule[-3ex]{0em}{8.0ex}} 
\def\st{\rule[-1.5ex]{0em}{4ex}}
\btabu{|l|c|c|}\hline
\multicolumn{1}{|c|}{\rw action}  
& ${ \dsp V(111) - V_{\rm fit}(\sqrt{3}) \over\dsp   V(2) - V(1)}$
& ${ \dsp V(221) - V(3) \over\dsp V(3) - V(1) }$  \\\hline\hline
\st Wilson           & 0.267(2)   &  0.099(3) \\[1ex]\hline
\st plaquette TI      & 0.091(10)  &  0.027(2) \\[1ex]\hline
\st mean-link TI & 0.050(5)   &  0.003(7) \\[1ex]\hline
\etabu
\caption[]{\label{tab:rotl}\small 
Comparisons of the normalized deviation from rotational invariance between the
Wilson action and the plaquette and mean-link tadpole-improved (TI) actions. The
ratios shown are zero in the case where rotational invariance holds, and on these 
measures the mean-link TI action is superior.
}
\enc
\etab

As mentioned in the introduction, there are many applications for
anisotropic lattices. It would be valuable to measure anisotropies for
a wider variety of lattice spacings and bare anisotropies, using the
methods we have investigated. It would also be very useful to
calculate perturbative formulae for the anisotropy and lattice spacing
as a function of the bare parameters of the action, $\beta$ and
$\chi_0$ \cite{buea,saea,drea2}.  Finally, it will be
necessary to repeat this tuning process for improved light quark actions. Some
of the requisite perturbative calculations have already been performed
\cite{groo}.

\newsection{Acknowledgements}
This work is supported in part by funds provided by the U.S.
Department of Energy (D.O.E.) under cooperative research agreement
\#DF-FC02-94ER40818. Part of the calculation was performed on the SP-2
at the Cornell Theory Center, which receives funding from Cornell
University, New York State, federal agencies, and corporate partners.
Calculations were carried out on the Hitachi SR2201 computers at the 
University of Cambridge High Performance Computing Facility and 
the Tokyo University Computer Centre.

\bibliography{refs}

\begin{thebibliography}{10}

\bibitem{MP}
C.~Morningstar and M.~Peardon.
\newblock {\em Phys. Rev.}, D56:4043, 1997.

\bibitem{MP0}
C.~Morningstar and M.~Peardon.
\newblock {\em Phys. Rev.}, D60:034509, 1999.

\bibitem{JKM}
K.~Juge, J.~Kuti, and C.~Morningstar.
\newblock {\em Nucl. Phys. Proc. Suppl.}, 63:326, 1998.

\bibitem{maea1}
T.~Manke et~al.
\newblock {\em Phys. Rev. Lett.}, 82:4396, 1999.

\bibitem{drea0}
I.T. Drummond et~al.
\newblock {\em hep-lat/9912041}.
\newblock To be published in Phys. Lett. B.

\bibitem{maea0}
T.~Manke et~al.
\newblock {\em hep-lat/9909038}.

\bibitem{alea}
M.~Alford, T.R. Klassen, and G.P. Lepage.
\newblock {\em Phys. Rev.}, D58:034503, 1998.

\bibitem{morn}
C.~Morningstar.
\newblock {\em Nucl.Phys.Proc.Suppl.}, 53:914--916, 1997.

\bibitem{buea}
G.~Burgers et~al.
\newblock {\em Nucl. Phys.}, B304:587, 1988.

\bibitem{saea}
S.~Sakai et~al.
\newblock {\em hep-lat/0002029}.

\bibitem{lepa}
G.P. Lepage.
\newblock {\em Nucl.Phys.Proc.Suppl.}, 60A:267--278, 1998.

\bibitem{drea1}
I.T. Drummond, S.~Duane, and R.R. Horgan.
\newblock {\em Nucl. Phys.}, B220 [FS8]:119--136, 1983.

\bibitem{tepe}
M.~Teper.
\newblock {\em Phys. Rev.}, D59:014512, 1999.

\bibitem{lues}
M.~L\"{u}scher.
\newblock {\em Nucl. Phys.}, B180:317, 1981.

\bibitem{ispa}
N.~Isgur and J~Paton.
\newblock {\em Phys. Rev.}, 31:2910--2929, 1985.

\bibitem{caea}
S.M. Catterall et~al.
\newblock {\em Phys. Lett.}, B321:246--253, 1994.

\bibitem{klas}
T.R. Klassen.
\newblock {\em Nucl. Phys.}, B533:557--575, 1998.

\bibitem{wegn}
F.J. Wegner.
\newblock {\em J. Phys. C}, 7:2098, 1974.

\bibitem{fira}
M.E. Fisher and M.~Randeria.
\newblock {\em Phys. Rev. Lett.}, 56:2332--2333, 1986.

\bibitem{luwe}
M.~L\"{u}scher and P.~Weisz.
\newblock {\em Comm. Math. Phys.}, 97:59, 1985.

\bibitem{caea0}
S.M. Catterall et~al.
\newblock {\em Phys. Lett.}, B300:393--399, 1993.

\bibitem{shtr0}
N.H. Shakespeare and H.D. Trottier.
\newblock {\em Phys. Rev.}, D59:014502, 1999.

\bibitem{shtr1}
N.H. Shakespeare and H.D. Trottier.
\newblock {\em Nucl. Phys. Proc. Suppl.}, 73:342--344, 1999.
\newblock Proceedings of LATTICE 98.

\bibitem{somm}
R.~Sommer.
\newblock {\em Nucl.Phys.Proc.Suppl.}, 60A:279--294, 1998.

\bibitem{luea}
M.~L\"{u}scher et~al.
\newblock {\em Nucl. Phys.}, B491:323, 1997.

\bibitem{drea2}
I.T. Drummond et~al.
\newblock Analytic two-loop calculation and simulation of anisotropic lattice
  parameters.
\newblock work in progress.

\bibitem{groo}
S.~Groote and J.~Shigemitsu.
\newblock {\em hep-lat/0001021}.

\end{thebibliography}
\bibliographystyle{unsrt}

\newpage
\befh
\bec
\epsfig{file=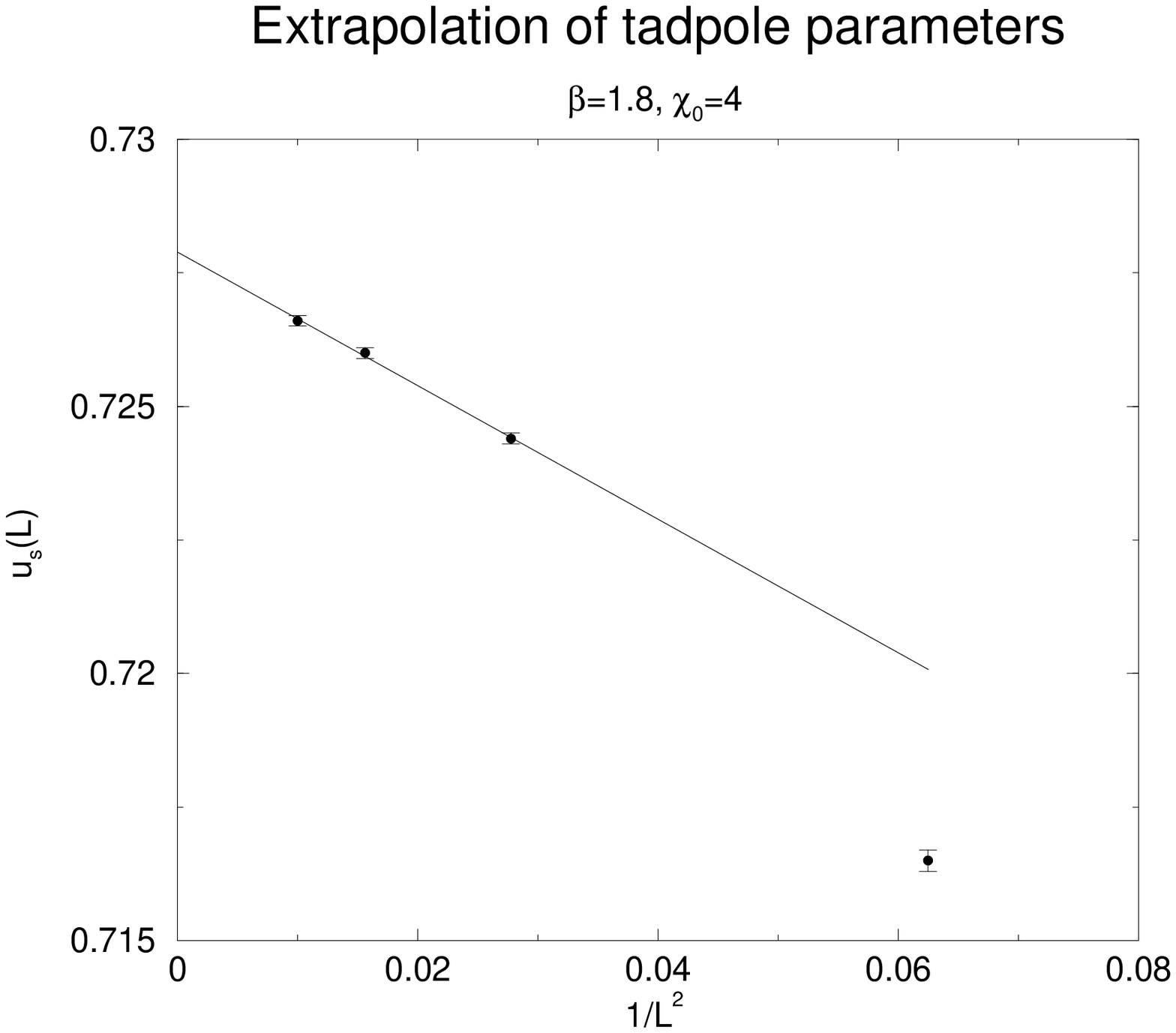,height=80mm}
\enc
\caption{\label{tad_s}\small
Fit of $u_s(L) = u_s(\infty)+A_s/L^2$ for $L=6,8,10$. 
}
\enf
\vskip 10mm
\befh
\bec
\epsfig{file=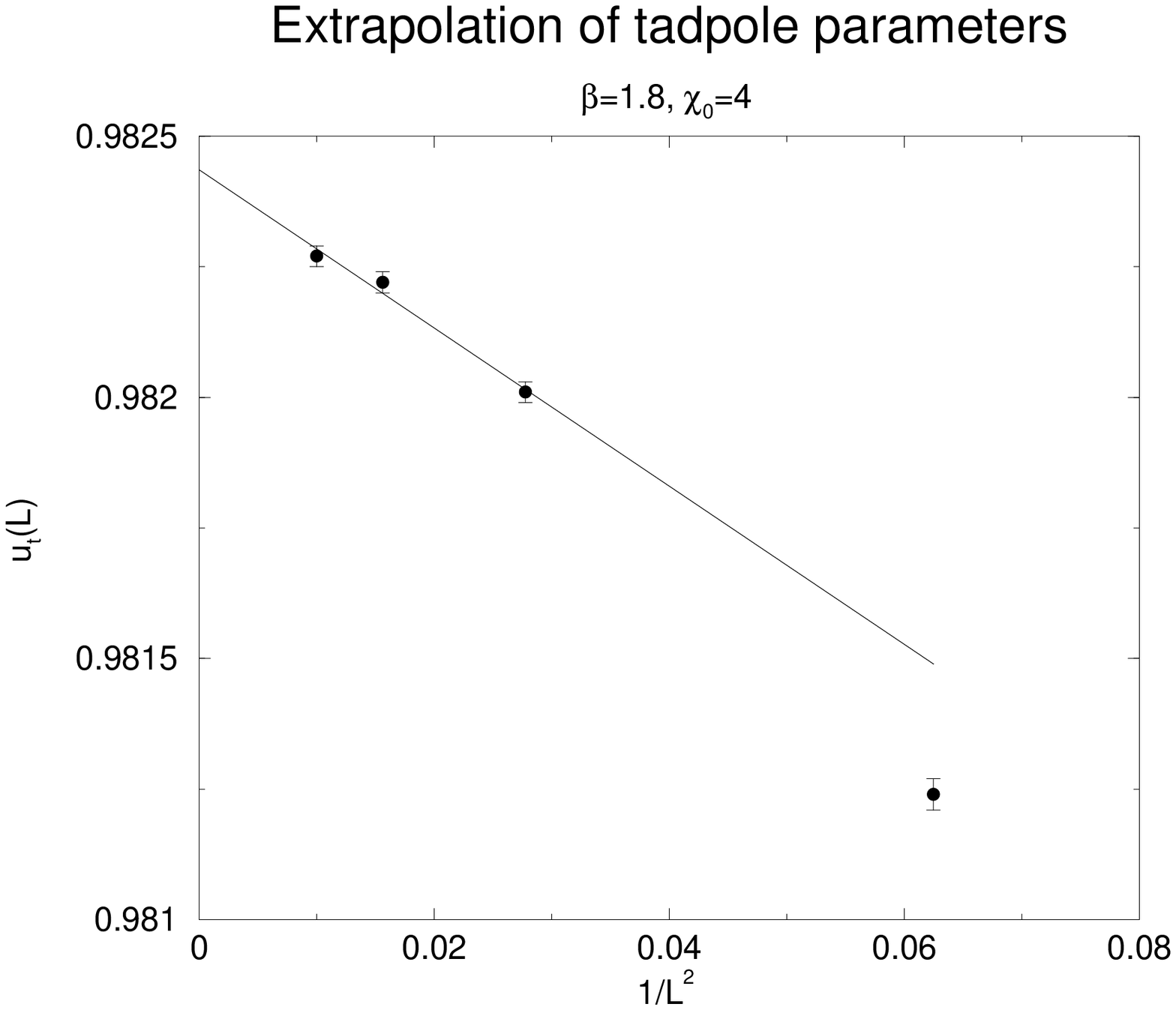,height=80mm}
\enc
\caption{\label{tad_t}\small
Fit of $u_t(L) = u_s(\infty)+A_t/L^2$ for $L=6,8,10$. 
}
\enf
\vskip 10mm
\befh
\bec
\epsfig{file=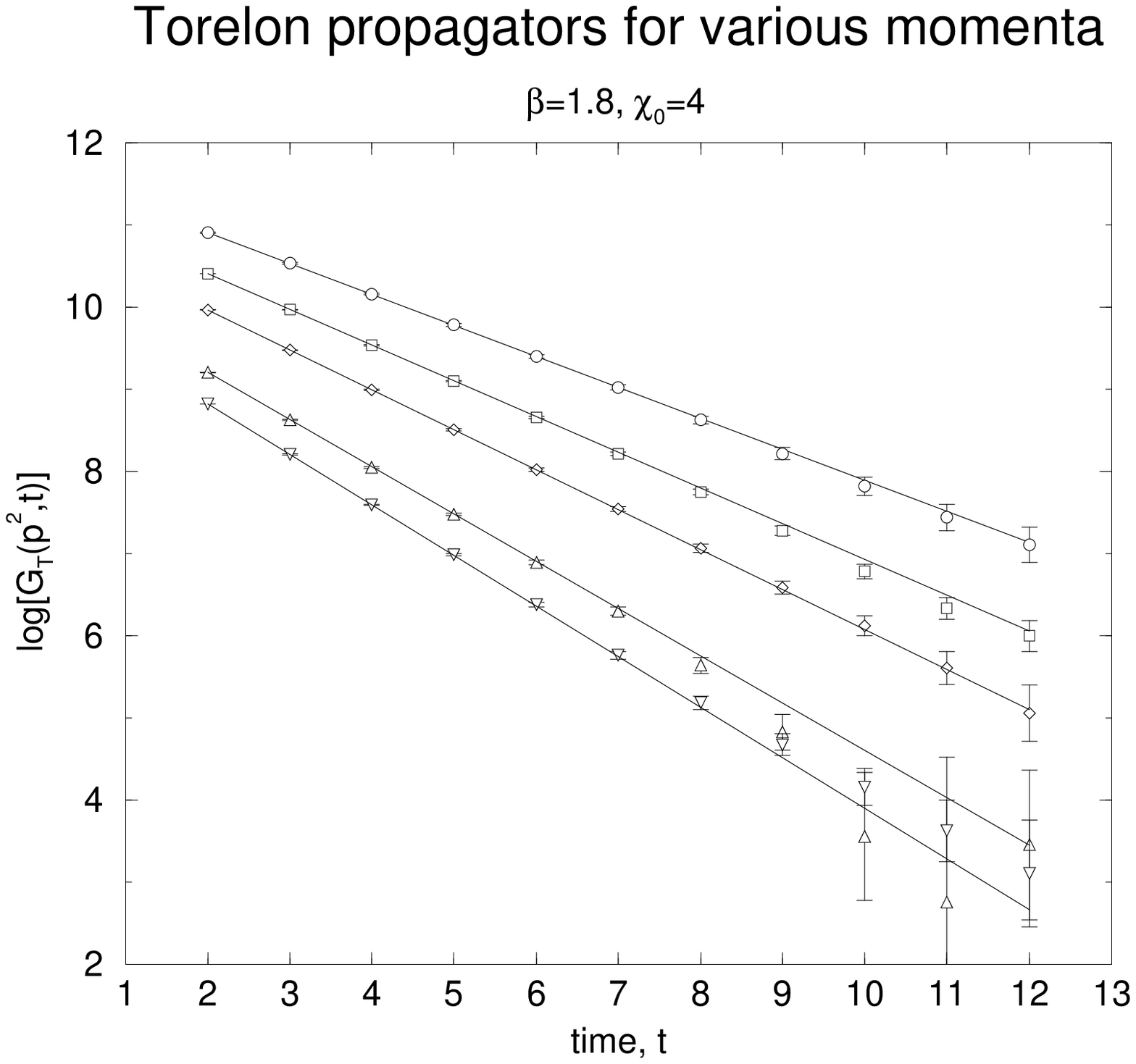,height=120mm}
\enc
\caption{\label{tor_fit}\small
Fit of the relativistic dispersion Eqn.~(\ref{disp}) to the torelon propagators for 
$\bp^2=\bn^2(\pi/4)^2$ for $\bn^2=0,1,2,4,5$ corresponding respectively to the curves from 
the top downwards. The fit has $\chi^2/N_{\rm df}=0.96$ with 
$(M_Ta_s)^2=1.85(4),~\chi_R=3.61(2)$.
}
\enf
\vskip 10mm
\befh
\bec
\epsfig{file=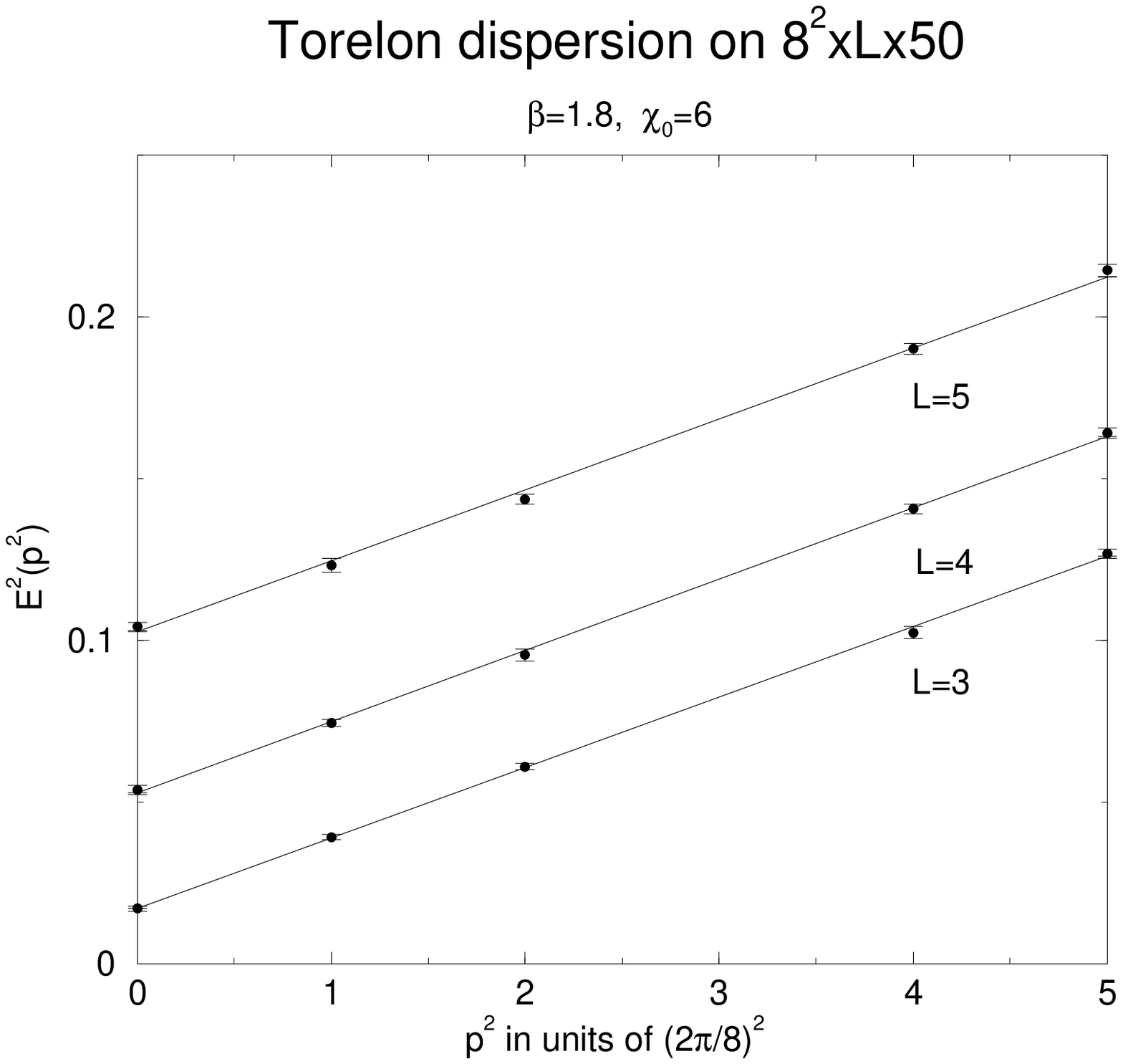,height=120mm}
\enc
\caption{\label{tor_disp}\small
Plots of $E(\bp^2)$ versus $\bp^2$ for $L=3,4,5$ on $8^2\times L\times50$ lattice. The 
lines show the fit to $E(\bp^2) = A+B\bp^2$.
}
\enf
\vskip 10mm
\befh
\bec
\epsfig{file=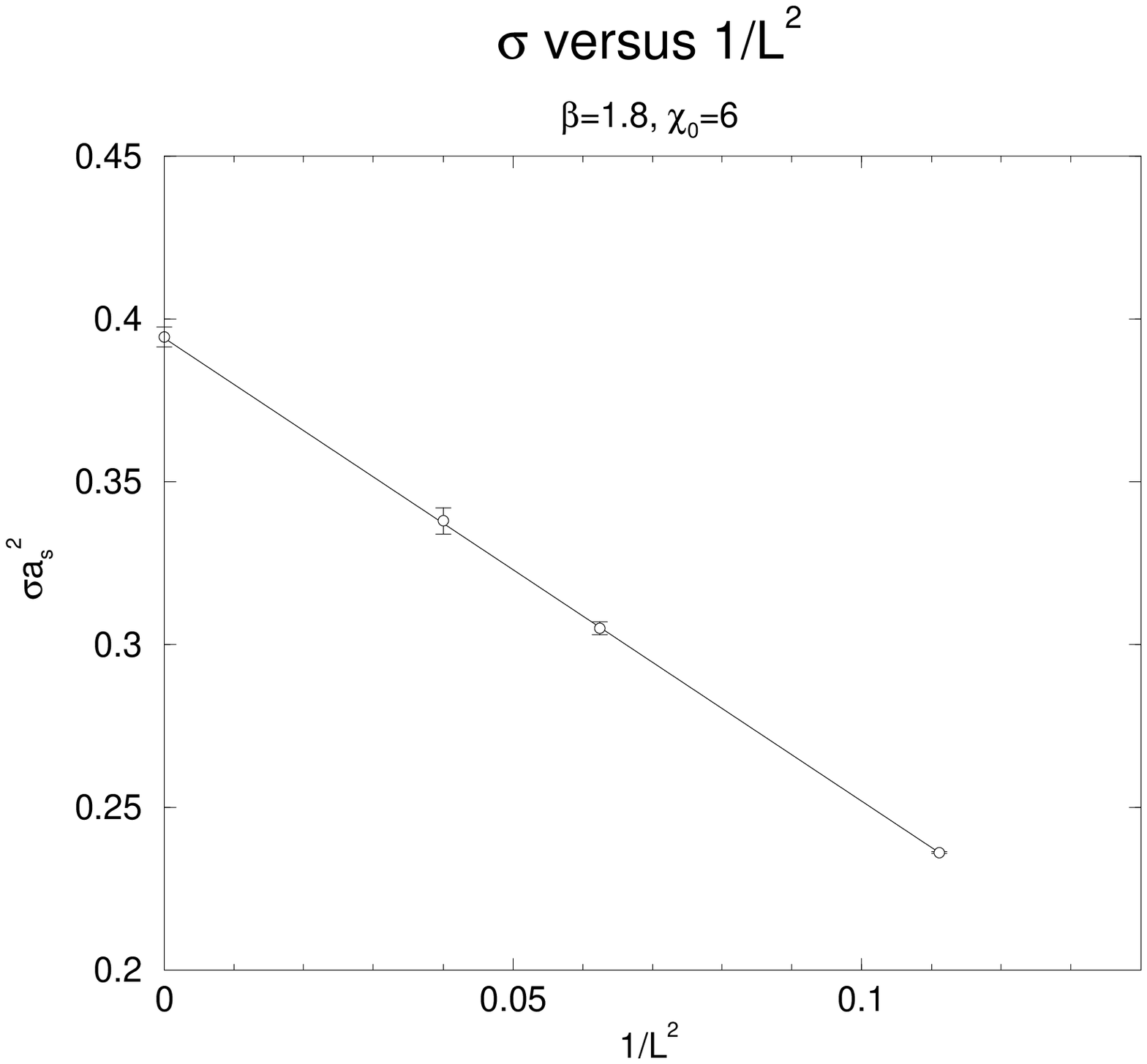,height=120mm}
\enc
\caption{\label{string_ten}\small
$\s(L)$ versus $1/L^2$ for $\gbL=1.8,~\chi_0=6$, with a linear fit giving
$\s a_s^2=0.394(3)$. 
}
\enf
\vskip 10mm
\befh
\bec
\epsfig{file=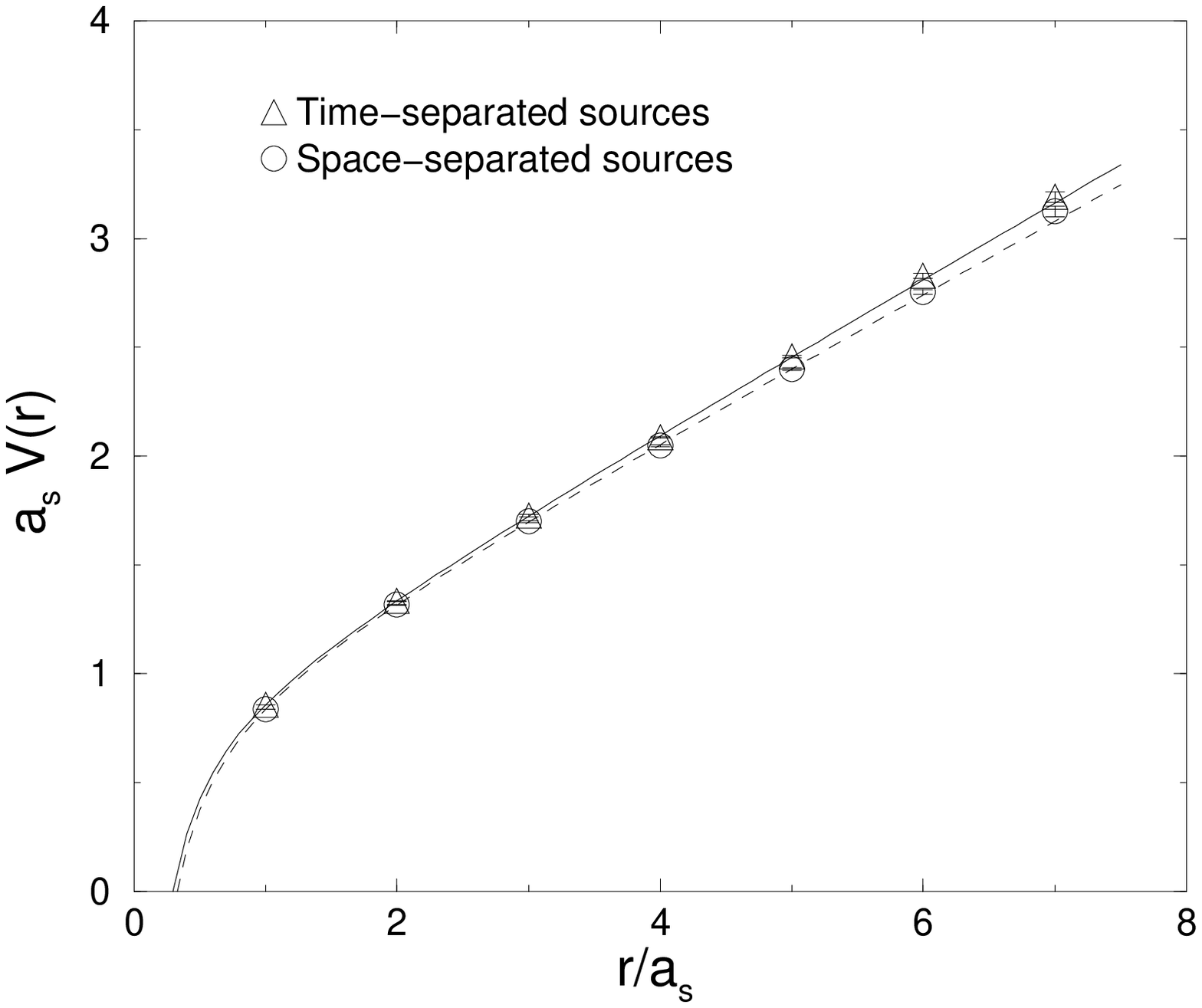,height=120mm}
\enc
\caption{\label{fig:plaq-sideways}\small
The potential at $\gbP=2.3, \chi_0=6$ between static sources separated along 
both spatial and temporal axes. 
}
\enf

\end{document}